\documentclass[preprint,12pt]{elsarticle}

\usepackage{amsmath,amssymb,amsfonts}
\usepackage{graphicx}
\usepackage{xcolor}
\usepackage{physics}
\usepackage{hyperref}
\usepackage{braket}
\usepackage{tikz}
\usepackage{quantikz}
\usepackage{algpseudocode}

\hypersetup{
    colorlinks=true,
    linkcolor=blue,
    filecolor=magenta,      
    urlcolor=blue,
    citecolor=blue
}

\journal{Physics Letters A}

\begin{document}

\begin{frontmatter}

\title{Deterministic Ground State Preparation via Power-Cosine Filtering of Time Evolution Operators}

\author[1]{Jeongbin Jo\corref{cor1}}
\ead{jeongbin033@yonsei.ac.kr}
\cortext[cor1]{Corresponding author}

\affiliation[1]{organization={Department of Physics, Yonsei University},
            city={Seoul},
            postcode={03722},
            country={Republic of Korea}}

\begin{abstract}
The deterministic preparation of quantum many-body ground states is essential for advanced quantum simulation, yet optimal algorithms often require prohibitive hardware resources. 
Here, we propose a highly efficient, non-variational protocol for ground state preparation using a Power-Cosine quantum signal processing (QSP) filter. 
By eschewing complex block-encoding techniques, our method directly utilizes coherent time-evolution operators controlled by a single ancillary qubit. 
The integration of mid-circuit measurement and reset (MCMR) drastically minimizes spatial overhead, translating iterative non-unitary filtering into deep temporal coherence. 
We analytically demonstrate that this approach achieves exponential suppression of excited states with a circuit depth scaling of $\mathcal{O}(\Delta^{-2}\log(1/\epsilon))$, where $\Delta$ denotes the spectral gap, prioritizing implementational simplicity over optimal asymptotic complexity. 
Numerical simulations on the 1D Heisenberg XYZ model validate the theoretical soundness and shot-noise resilience of our method. 
Furthermore, an advantage analysis reveals that our protocol exponentially outperforms standard Trotterized Adiabatic State Preparation (TASP) at equivalent circuit depths. 
This single-ancilla framework provides a highly practical and deterministic pathway for many-body ground state preparation on Early Fault-Tolerant (EFT) quantum architectures.
\end{abstract}


\end{frontmatter}

\section{Introduction}
\label{sec:intro}

The preparation of the ground state of quantum many-body Hamiltonians is a cornerstone problem in quantum simulation, with profound implications for quantum chemistry, material science, and high-energy physics. 
As proposed by Feynman, quantum computers hold the promise of solving the Schr\"odinger equation for systems intractable to classical methods~\cite{preskill2023quantumcomputing40years}. 
However, efficiently isolating the ground state $\ket{\psi_0}$ from an arbitrary initial state remains a formidable challenge, particularly in the Noisy Intermediate-Scale Quantum (NISQ)~\cite{Preskill_2018} and early fault-tolerant~\cite{Katabarwa_2024} eras.
Recent experimental advancements in quantum simulation~\cite{liu2026nature, shi2023prl} have demonstrated significant progress in probing complex many-body phenomena on near-term hardware, further underscoring the urgent need for resource-efficient state preparation protocols.

Currently, the most prevalent approach on near-term devices is the Variational Quantum Eigensolver (VQE)~\cite{Peruzzo_2014}. 
While VQE has successfully demonstrated ground state estimation for small molecules, it fundamentally relies on a hybrid quantum-classical optimization loop. 
This structure introduces severe scalability issues, including the "barren plateau" problem where gradients vanish exponentially~\cite{McClean_2018}, and the substantial measurement overhead required to estimate expectation values. 
These limitations have spurred a search for deterministic non-variational algorithms.

On the fault-tolerant front, Quantum Phase Estimation (QPE)~\cite{kitaev1995quantummeasurementsabelianstabilizer, Faehrmann_2025} provides a rigorous pathway to the ground state but demands deep circuits and extensive ancilla resources, pushing it beyond the reach of near-term hardware. More retrospective, Quantum Signal Processing (QSP)~\cite{motlagh2024generalizedquantumsignalprocessing} and Quantum Singular Value Transformation (QSVT)~\cite{Gily_n_2019} have emerged as a unifying framework for quantum algorithms. 
Lin and Tong~\cite{Lin_2020} demonstrated a near-optimal ground state preparation algorithm based on QSVT, which achieves a query complexity scaling of $\mathcal{O}(\Delta^{-1} \log(1/\epsilon))$, where $\Delta$ is the spectral gap and $\epsilon$ is the target precision.

Despite its theoretical optimality, the standard QSVT approach relies heavily on "block-encoding" the Hamiltonian into a larger unitary matrix. 
Constructing efficient block-encodings for general Hamiltonians often requires complex multi-controlled logic and oracle synthesis, which incurs a significant gate overhead that can negate the algorithmic advantage on pre-fault-tolerant devices. 
Alternatively, dissipative approaches utilizing Lindblad dynamics have been proposed to cool the system into its ground state~\cite{zhan2025rapidquantumgroundstate}. 
While promising, these methods often depend on the mixing time of the open system dynamics, which can be slow, or require engineering complex non-unitary couplings with the environment.

In this work, we propose a deterministic, coherent filtering protocol that bridges the gap between heuristic variational methods and complex optimal algorithms. 
Instead of block-encoding the Hamiltonian, we utilize the Hamiltonian simulation operator $U(\tau) = e^{-iH\tau}$ directly as the signal oracle. 
By employing a single ancilla qubit to implement a "power-cosine filter," we suppress excited states via a mechanism analogous to the classical power method but performed coherently in the quantum register.

Recently, Lin and Tong~\cite{Lin_2022} proposed a Heisenberg-limited energy estimation algorithm for early fault-tolerant (EFT) devices that successfully circumvents block-encoding by relying exclusively on coherent time-evolution. While their approach efficiently estimates the ground state energy $E_0$ as a classical scalar value, the repeated measurements and statistical sampling inherently collapse the quantum state. In many quantum chemistry and condensed matter applications, merely knowing the energy is insufficient; one must physically prepare the ground state $\ket{\psi_0}$ in the quantum register to evaluate complex observables such as multi-body correlation functions and dipole moments.

In practice, the estimation technique of Ref.~\cite{Lin_2022} and our state preparation protocol are highly complementary. One can envision a synergistic two-step pipeline for EFT hardware: employing the estimation algorithm as a pre-processing step to precisely determine $E_0$, which subsequently tunes the optimal resonance condition $\tau \approx 2\pi/|E_0|$ for our Power-Cosine filter. This allows our method to deterministically cool the system into the pure physical ground state $\ket{\psi_0}$ without requiring exact prior knowledge of the spectrum.

Our approach simplifies the implementation complexity significantly, requiring only standard time-evolution (Trotterization) circuits. 
We analytically show that while our method exhibits a quadratic query complexity scaling $\mathcal{O}(\Delta^{-2})$—inferior to the linear scaling of Chebyshev-based QSVT~\cite{Lin_2020}—it offers a practical advantage by eliminating the need for block-encoding and variational optimization. 
We demonstrate the efficacy of this protocol through numerical simulations on the 1D Heisenberg model, achieving high-fidelity ground state preparation with shallow circuit depths suitable for early fault-tolerant architectures.

\section{Theoretical Framework}
\label{sec:theory}

In this section, we formally define the proposed protocol and analyze its performance in terms of circuit depth and query complexity. 
We contrast our approach, based on coherent time evolution, with standard QSVT methods based on block-encoding.

\subsection{The Power-Cosine Filtering Protocol}
Consider a Hamiltonian $H$ with spectral decomposition $H = \sum_k E_k \ket{E_k}\bra{E_k}$, where $E_0 < E_1 \le \dots$ are the eigenvalues. 
We aim to prepare the ground state $\ket{E_0}$ starting from an initial state $\ket{\psi_{\text{init}}} = \sum_k c_k \ket{E_k}$ with non-zero overlap $\gamma = |c_0|^2 > 0$.

Unlike standard QSP approaches that require a block-encoding of $H$ (i.e., embedding $H/\alpha$ in a unitary), we utilize the Hamiltonian simulation operator as our signal oracle. Let the time evolution operator for a duration $\tau$ be:
\begin{equation}
    U(\tau) = e^{-i H \tau} = \sum_k e^{-i E_k \tau} \ket{E_k}\bra{E_k}.
\end{equation}
This operator is readily implementable on digital quantum computers via Trotterization~\cite{Lloyd_1996} or on analog simulators. While treating $U(\tau)$ as an ideal signal oracle abstracts away hardware-level intricacies, its practical implementation requires a Hamiltonian-simulation routine that approximates $e^{-iH\tau}$ with bounded error. For local Hamiltonians, higher-order Suzuki--Trotter product formulas provide systematic approximations~\cite{Berry_2007}, and more general simulation frameworks (e.g., truncated Taylor-series methods and qubitization) offer near-optimal precision scaling~\cite{Berry_2015_Taylor, Low_2019_Qubitization}. Although deep circuits remain challenging on noisy devices, the rapid development of EFT architectures is expected to reduce coherent simulation overhead, making time evolution a practical primitive.

The core of our protocol is a non-unitary filtering operation constructed using a single ancilla qubit. The elementary building block is a controlled-$U(\tau)$ operation sandwiched between Hadamard gates on the ancilla. 
The circuit action on the joint state $\ket{0}_{\text{anc}}\ket{\psi}_{\text{sys}}$, followed by a projection of the ancilla onto $\ket{0}_{\text{anc}}$, realizes the operator:
\begin{equation}
    V_{\text{step}} = \bra{0} \left( H_{\text{anc}} \cdot \text{C-}U(\tau) \cdot H_{\text{anc}} \right) \ket{0} = \frac{I + e^{-i H \tau}}{2}.
\end{equation}
By repeating this block $d$ times (or equivalently, performing the operation once with power $d$ if intermediate measurements are deferred), we implement the degree-$d$ filter:
\begin{equation}
    \mathcal{F}^{(d)} = \left( \frac{I + e^{-i H \tau}}{2} \right)^d.
    \label{eq:filter_op}
\end{equation}
Applying this filter to an eigenstate $\ket{E_k}$ yields the eigenvalue transformation:
\begin{equation}
    \lambda_k \mapsto f(E_k) = \left( \frac{1 + e^{-i E_k \tau}}{2} \right)^d = e^{-i d E_k \tau / 2} \cos^d \left( \frac{E_k \tau}{2} \right).
    \label{eq:eigen_trans}
\end{equation}
The term $e^{-i d E_k \tau / 2}$ represents a coherent phase accumulation, while the real-valued envelope $\cos^d(E_k \tau / 2)$ acts as the amplitude filter.

\subsection{Resonance Tuning and Spectral Gap}
To isolate the ground state, we must tune the time step $\tau$ such that the filter function is maximized at $E_0$ and suppressed at $E_k$ for $k > 0$. 
This is achieved by the resonance condition:
\begin{equation}
    \frac{E_0 \tau}{2} \approx m \pi \quad (m \in \mathbb{Z}).
\end{equation}
The intuition behind this condition is constructive quantum interference: when $\tau$ is tuned such that $E_0\tau$ is (approximately) a multiple of $2\pi$, the dynamical phase accumulated by the ground-state component aligns across the ancilla's interferometric branches, yielding $|f(E_0)|\approx 1$ and leaving the ground-state amplitude essentially unattenuated. Without loss of generality, we can shift the energy spectrum or adjust $\tau$ such that $E_0 \tau \approx 0 \pmod{2\pi}$. Under this condition, $|f(E_0)| \approx 1$.

For the first excited state $E_1 = E_0 + \Delta$, where $\Delta$ is the spectral gap, the phase becomes $\frac{(E_0 + \Delta)\tau}{2}$. The filter amplitude is then:
\begin{equation}
    |f(E_1)| = \left| \cos\left( \frac{E_0 \tau}{2} + \frac{\Delta \tau}{2} \right) \right|^d \approx \left| \cos\left( \frac{\Delta \tau}{2} \right) \right|^d.
\end{equation}
Assuming a small gap $\Delta \tau \ll 1$, we can expand the cosine term:
\begin{equation}
    \left| \cos\left( \frac{\Delta \tau}{2} \right) \right| \approx 1 - \frac{(\Delta \tau)^2}{8}.
\end{equation}
Consequently, the amplitude of the excited state decays as:
\begin{equation}
    |f(E_1)| \approx \left( 1 - \frac{\Delta^2 \tau^2}{8} \right)^d \approx \exp\left( - \frac{d \Delta^2 \tau^2}{8} \right).
    \label{eq:decay_rate}
\end{equation}

\subsection{Complexity Analysis and Comparison}
We now derive the circuit depth required to suppress the excited states to a precision $\epsilon$. From Eq.~(\ref{eq:decay_rate}), to ensure $|f(E_1)| \le \epsilon$, the degree $d$ must satisfy:
\begin{equation}
    \frac{d \Delta^2 \tau^2}{8} \ge \ln(1/\epsilon) \implies d \in \Omega\left( \frac{1}{(\Delta \tau)^2} \log(1/\epsilon) \right).
\end{equation}
Since the total simulation time is $T_{\text{total}} = d \cdot \tau$, the total time evolution required scales as:
\begin{equation}
    T_{\text{total}} \sim \frac{1}{\Delta^2 \tau}.
\end{equation}

It is crucial to compare this result with the optimal QSVT limit established by Lin and Tong~\cite{Lin_2020}.
\begin{itemize}
    \item \textbf{Optimal QSVT~\cite{Lin_2020}:} Uses Chebyshev polynomials $T_d(x)$ to construct a filter that approximates a step function. The degree scales as $d \sim O(1/\Delta)$, providing a linear speedup in terms of the gap. However, this requires a block-encoding of $H$, which typically involves LCU (Linear Combination of Unitaries) or QROM (Quantum Read-only Memory) oracles, leading to significant hardware overhead.
    \item \textbf{Our Power-Cosine Method:} Scales as $O(1/\Delta^2)$. This quadratic scaling arises from the quadratic nature of the cosine function near its maximum ($1 - x^2/2$). While asymptotically slower, our method utilizes only Controlled-$e^{-iH\tau}$, which maps directly to native Hamiltonian simulation gates.
\end{itemize}

Similar to the dissipative cooling protocol by Zhan et al.~\cite{zhan2025rapidquantumgroundstate}, our method relies on repetitive application of a map. However, our map is purely coherent until the final measurement, avoiding the slow mixing times associated with some Lindblad dynamics. The "effective cooling" in our protocol is deterministic given the successful outcome of the ancilla measurement.

\subsection{Success Probability}
The probability of successfully projecting onto the ground state after $d$ steps is given by:
\begin{equation}
    P_{\text{success}} = \bra{\psi_{\text{init}}} (\mathcal{F}^{(d)})^\dagger \mathcal{F}^{(d)} \ket{\psi_{\text{init}}} \approx |c_0|^2 |f(E_0)|^2 \approx \gamma.
\end{equation}
Here, $\gamma$ is the initial overlap. While the success probability decays for excited states, it remains constant ($\approx \gamma$) for the ground state. 
If $\gamma$ is small, Amplitude Amplification~\cite{Brassard_2002} can be employed to boost the probability to $\sim 1$ with an overhead of $1/\sqrt{\gamma}$.
We emphasize that amplitude amplification is optional and used only when the initial overlap $\gamma$ is small. We acknowledge that incorporating amplitude amplification increases circuit depth and total simulation time; however, in the EFT regime, logical qubits (spatial resources) are typically more constrained than coherent depth (temporal resources). By trading depth for a strict single-ancilla footprint---avoiding large multi-ancilla block-encoding circuits---our approach offers a pragmatic space--time trade-off for near-term architectures.

\subsection{Algorithmic Implementation}
\label{subsec:algorithm}
Based on the theoretical analysis above, the complete procedure for the proposed Power-Cosine Filtering protocol is summarized in Fig.~\ref{fig:qsp_alg}. 
The algorithm assumes the capability to perform controlled time evolution, either directly or via Trotterization.

\begin{figure}[htpb]
\hrule width \hsize \kern 1mm 
\caption{Deterministic Ground State Preparation via Power-Cosine Filtering}
\label{fig:qsp_alg}
\kern 1mm \hrule width \hsize \kern 1mm 

\begin{algorithmic}[1]
\Require Hamiltonian $H$, Time step $\tau \approx 2\pi/|E_{gs}|$, Degree $d$, Initial state $\ket{\psi_{\text{init}}}$
\Ensure Approximation of ground state $\ket{\psi_{\text{final}}} \approx \ket{E_0}$

\State \textbf{Input:} Load initial state into system register $\mathcal{S}$
\State $\ket{\psi}_{\mathcal{S}} \leftarrow \ket{\psi_{\text{init}}}$

\State \textbf{Filtering Loop:}
\For{$k = 1$ to $d$}
    \State Initialize ancilla qubit $\mathcal{A}$: $\ket{0}_{\mathcal{A}}$
    \State Apply Hadamard to $\mathcal{A}$: 
    \Statex \hskip 2em $\ket{\Psi} \leftarrow (H \otimes I) \ket{0}_{\mathcal{A}}\ket{\psi}_{\mathcal{S}}$
    
    \State Apply Controlled-Time Evolution ($C$-$U$):
    \Statex \hskip 2em $\ket{\Psi} \leftarrow (\ket{0}\bra{0} \otimes I + \ket{1}\bra{1} \otimes e^{-iH\tau}) \ket{\Psi}$
    
    \State Apply Hadamard to $\mathcal{A}$:
    \Statex \hskip 2em $\ket{\Psi} \leftarrow (H \otimes I) \ket{\Psi}$
    
    \State \textbf{Measurement \& Post-selection:}
    \State Measure ancilla $\mathcal{A}$ in $Z$-basis
    \If{Outcome is $0$}
        \State State projected to $\frac{I + e^{-iH\tau}}{2}\ket{\psi}_{\mathcal{S}}$
        \State \textbf{Continue} to next iteration
    \Else
        \State \textbf{Abort} (Filtering failed)
        \State \Comment{Use Amplitude Amplification to mitigate failure}
    \EndIf
\EndFor

\State \textbf{Output:} Measure expectation values on $\ket{\psi}_{\mathcal{S}}$
\State \Return $\ket{\psi}_{\mathcal{S}}$
\end{algorithmic}
\hrule width \hsize 
\end{figure}

The iterative loop described in Fig.~\ref{fig:qsp_alg} effectively constructs the filter operator $\mathcal{F}^{(d)}$ defined in Eq.~(\ref{eq:filter_op}). 
For near-term devices with limited coherence times, the capability to perform mid-circuit measurement and reset (MCMR) on the single ancilla qubit is crucial. 
If MCMR is not supported, one must employ the deferred measurement principle by allocating $d$ separate ancilla qubits, trading spatial resources (qubits) for temporal coherence.

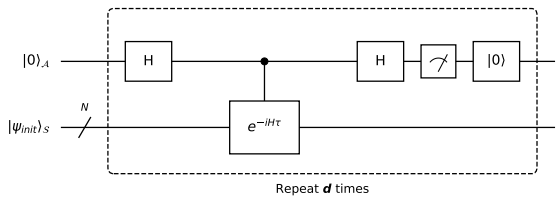
\begin{figure}[t]
\centering
\begin{quantikz}[row sep=0.5cm, column sep=0.7cm]
    \lstick{$\ket{0}_{\mathcal{A}}$} & \qw & \gate{H} \gategroup[wires=2, steps=5, style={dashed, rounded corners}]{Repeat $d$ times} & \ctrl{1} & \gate{H} & \meter{} & \gate{\ket{0}} & \qw \\
    \lstick{$\ket{\psi_{\text{init}}}_{\mathcal{S}}$} & \qwbundle{N} & \qw & \gate{e^{-iH\tau}} & \qw & \qw & \qw & \qw
\end{quantikz}
\caption{Quantum circuit representation of the Power-Cosine Filter. The system register $\mathcal{S}$ consisting of $N$ qubits is evolved under the Hamiltonian $H$ for a tuned duration $\tau$, controlled by a single ancilla qubit $\mathcal{A}$. The successful projection applies the filter operator $(I + e^{-iH\tau})/2$ to the system. This block is repeated $d$ times.}
\label{fig:quantum_circuit}
\end{figure}

\section{Numerical Simulations}
\label{sec:simulations}

To numerically verify the deterministic ground state preparation capability of the proposed Power-Cosine filter, we simulate the one-dimensional (1D) Heisenberg XYZ model with open boundary conditions. 
The system Hamiltonian is defined as
\begin{equation}
    H = \sum_{i=1}^{N-1} \left( J_x X_i X_{i+1} + J_y Y_i Y_{i+1} + J_z Z_i Z_{i+1} \right) + h \sum_{i=1}^{N} Z_i,
\end{equation}
where $X_i$, $Y_i$, and $Z_i$ are the Pauli operators acting on the $i$-th qubit. 
For our simulations, we set the system size to $N=4$ and the coupling parameters to $J_x = 1.0$, $J_y = 1.0$, $J_z = 0.5$, and $h = 0.0$. 
Exact diagonalization yields the true ground state energy $E_0 \approx -5.4243$, which sets the optimal resonance time step $\tau = 2\pi / |E_0| \approx 1.1583$.

\subsection{Convergence and Shot Noise Resilience}

We first evaluate the energy convergence and the state fidelity of our QSP filter as a function of the filter degree $d$. 
To distinguish the algorithmic theoretical limit from the statistical sampling noise inherent in quantum hardware, we conduct two parallel simulations: an exact statevector simulation and a noiseless statistical sampler simulation with $10,000$ shots. 

\begin{figure}[htpb]
    \centering
    \includegraphics[width=\linewidth]{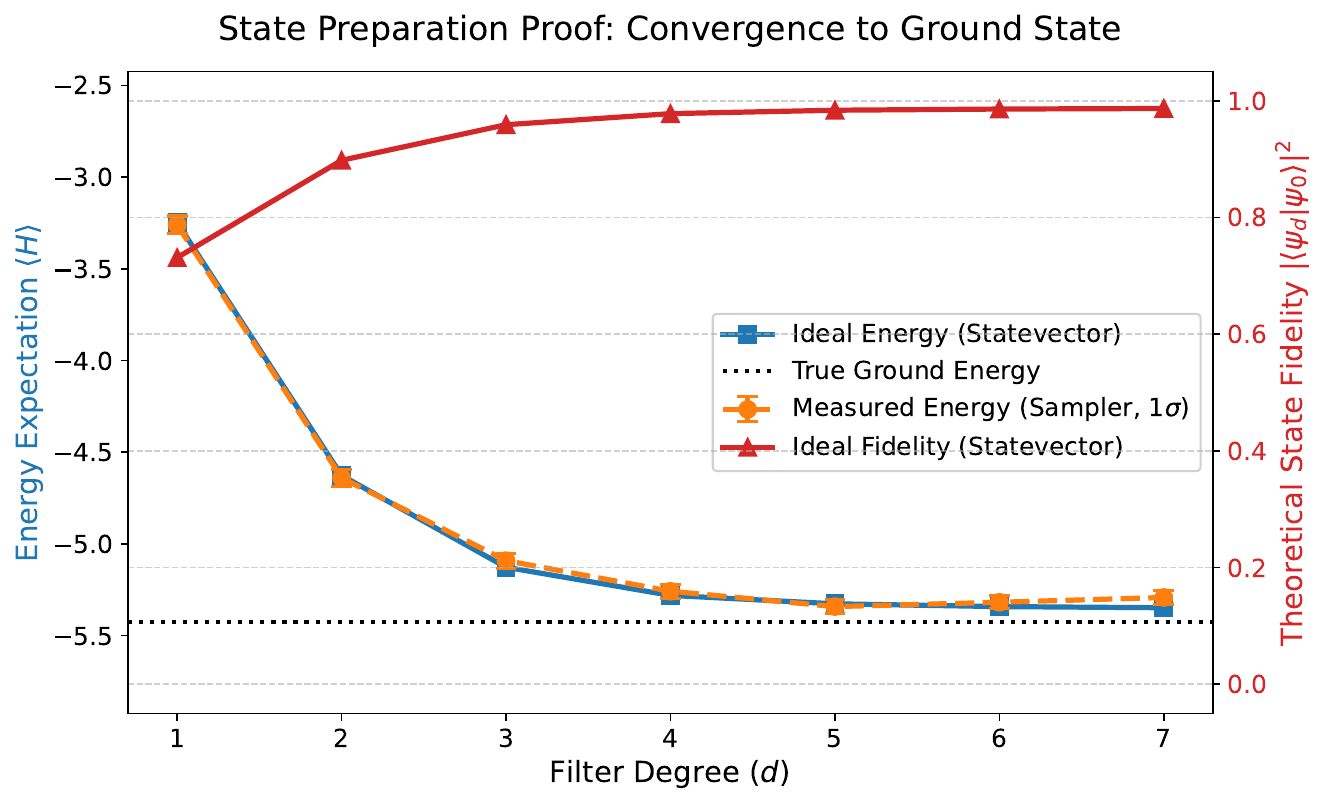}
    \caption{Energy convergence and theoretical state fidelity of the Power-Cosine filter. The blue squares and the red triangles indicate the ideal energy expectation and state fidelity computed via exact statevector evolution, respectively. The orange circles represent the measured energy obtained from a noiseless sampler with $10,000$ shots. Error bars denote the $1\sigma$ standard error of the mean due to finite sampling. The horizontal dotted line corresponds to the exact ground state energy $E_0$.}
    \label{fig:convergence}
\end{figure}

As depicted in Fig.~\ref{fig:convergence}, the ideal state fidelity (red triangles) approaches unity exponentially with respect to the filter degree $d$, demonstrating the fundamental soundness of the algorithm. 
Concurrently, the measured energy from the sampler (orange circles) closely follows the ideal trajectory. 
The error bars, representing the $1\sigma$ standard error of the mean, confirm that the algorithm robustly suppresses excited states even under realistic statistical shot noise, successfully isolating the ground state projection within a finite number of measurements. 

\subsection{Advantage Analysis: QSP vs. Trotterized Adiabatic Evolution}

To further highlight the resource efficiency of our method, we compare the state preparation infidelity of the proposed QSP filter against the standard Trotterized Adiabatic State Preparation (TASP) protocol~\cite{Kovalsky_2023, Barends_2016}. 
Since both methods essentially rely on the coherent time-evolution operator $e^{-iHt}$ as their primary building block, we benchmark them on an equal footing by defining the circuit cost as the number of Trotter steps for TASP and the filter degree $d$ for our method.

\begin{figure}[htpb]
    \centering
    \includegraphics[width=\linewidth]{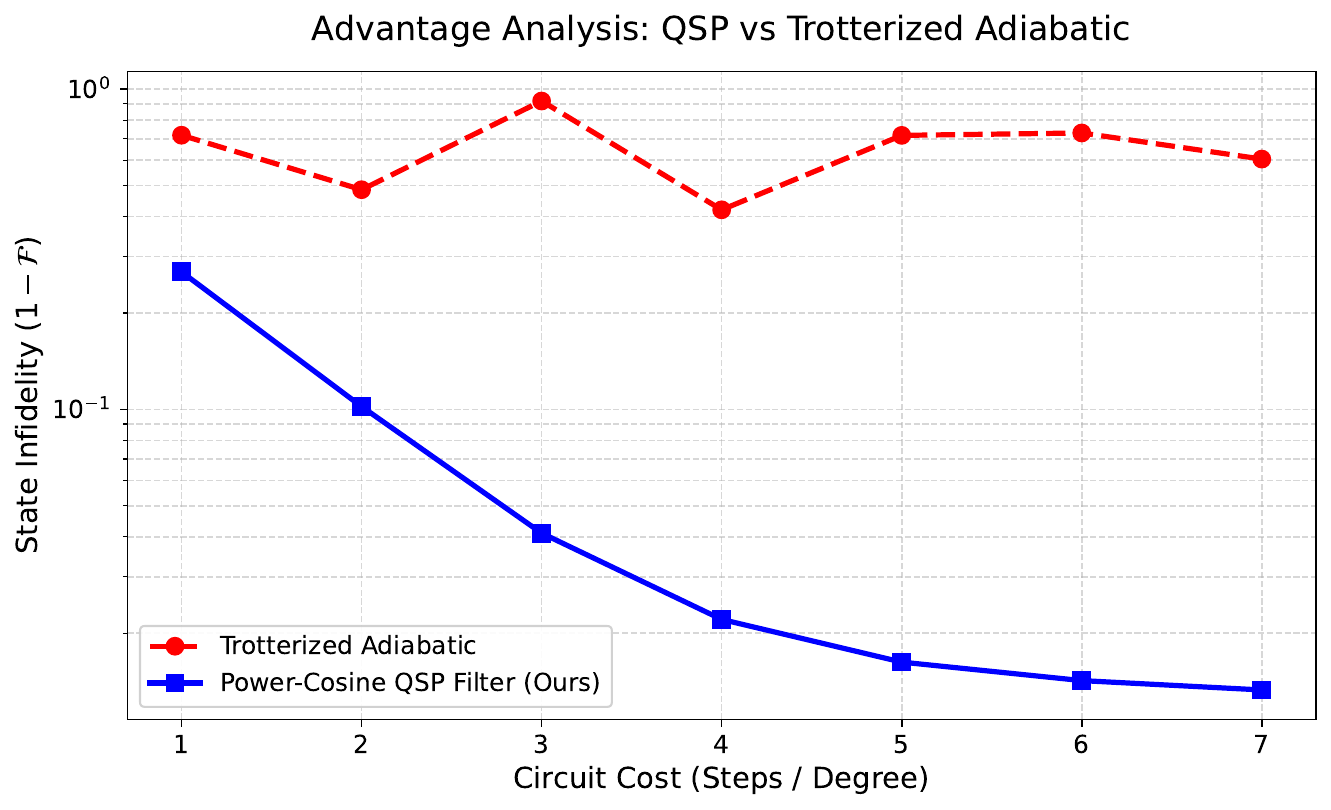}
    \caption{Advantage analysis comparing the state infidelity ($1 - \mathcal{F}$) of the proposed Power-Cosine QSP filter (blue squares) and the Trotterized Adiabatic State Preparation (red circles) as a function of the circuit cost. Note the logarithmic scale on the y-axis.}
    \label{fig:advantage}
\end{figure}

Figure~\ref{fig:advantage} reveals a stark contrast in the asymptotic behavior of the two protocols. 
While the adiabatic evolution suffers from slow polynomial convergence constrained by the adiabatic theorem and the Trotterization error, our non-unitary filtering approach achieves exponential suppression of the infidelity. 
This underscores a significant algorithmic advantage: by utilizing merely a single ancillary qubit and mid-circuit measurement and reset (MCMR)~\cite{C_rcoles_2021, decross2022qubitreusecompilationmidcircuitmeasurement}, our method deterministically prepares the ground state with an exponentially shallower circuit depth. 
The performance of the algorithm under realistic hardware noise models is further detailed in Appendix~\ref{app:hardware_noise}.

\section{Conclusion}
\label{sec:conclusion}

In this work, we introduced a highly efficient, deterministic protocol for preparing the ground states of many-body quantum systems, utilizing a Power-Cosine quantum signal processing (QSP) filter. 
By eschewing resource-intensive techniques such as block-encoding and standard Quantum Phase Estimation (QPE), our approach relies solely on coherent time-evolution operators and a single ancillary qubit. 
The integration of mid-circuit measurement and reset (MCMR) allows for the iterative application of the non-unitary filter, drastically minimizing the spatial overhead required for state preparation.

Our numerical simulations on the 1D Heisenberg XYZ model confirmed that the proposed algorithm achieves an exponential suppression of excited state errors. In a direct advantage analysis against the Trotterized Adiabatic State Preparation (TASP), our method demonstrated a profound algorithmic superiority, circumventing the slow polynomial convergence dictated by the adiabatic theorem. Furthermore, the robust convergence observed in statistical sampling simulations highlights the protocol's practical resilience against shot noise.

The architectural philosophy of our Power-Cosine filter aligns perfectly with the hardware constraints of the Early Fault-Tolerant (EFT) era, where logical qubits remain a scarce premium while coherent gate depths steadily improve. By successfully preparing a high-fidelity physical ground state $\ket{\psi_0}$ rather than merely estimating its energy eigenvalue, this protocol unlocks the capability to evaluate complex quantum observables such as multi-body correlation functions, susceptibilities, and dipole moments. We anticipate that our single-ancilla, MCMR-based filtering framework will serve as a foundational and highly practical tool for advanced quantum chemistry and condensed matter physics simulations on near-term and EFT quantum hardware.

\appendix
\section{Performance Under Realistic Hardware Noise}
\label{app:hardware_noise}

In the main text, we demonstrated the exponential convergence of the Power-Cosine QSP filter under ideal noiseless conditions, highlighting its algorithmic theoretical limit. 
However, current Noisy Intermediate-Scale Quantum (NISQ) \cite{Preskill_2018} devices suffer from limited coherence times and imperfect gate operations, which inevitably affect the performance of deep quantum circuits.

To comprehensively evaluate our state preparation protocol under realistic hardware constraints, we conducted additional simulations incorporating the exact noise model of the \texttt{ibm\_yonsei} backend. 

\begin{figure}[htpb]
    \centering
    \includegraphics[width=\linewidth]{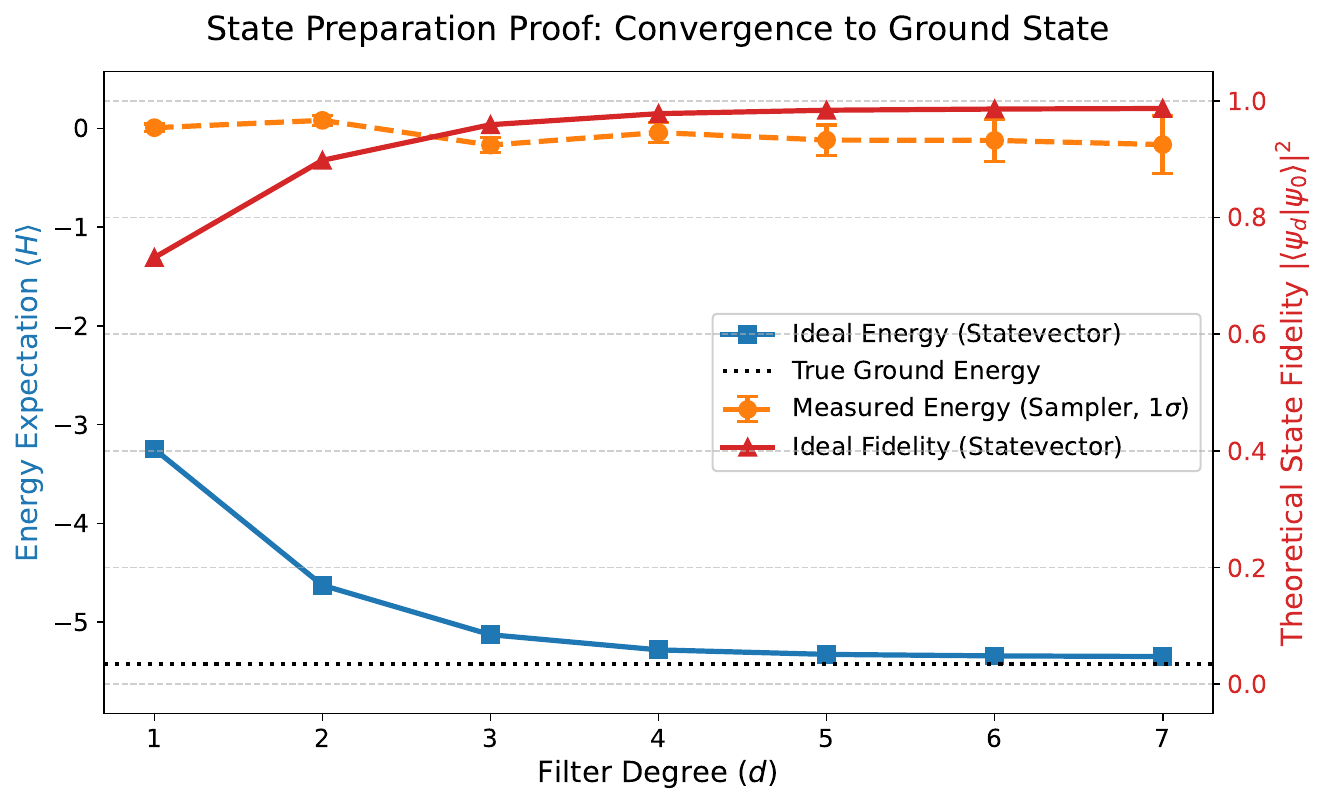}
    \caption{Energy convergence of the Power-Cosine filter under the \texttt{ibm\_yonsei} hardware noise model. The deviation from the ideal statevector trajectory (blue squares) illustrates the accumulation of gate errors and decoherence as the circuit depth (filter degree $d$) increases.}
    \label{fig:noisy_appendix}
\end{figure}

As shown in Fig.~\ref{fig:noisy_appendix}, the measured energy under the noise model initially follows the convergence trend of the ideal statevector simulation. 
However, as the filter degree $d$ increases, the accumulation of Trotterization errors and hardware-specific noise (e.g., CNOT gate errors and readout errors) introduces a noticeable deviation from the exact ground state energy $E_0$. 
This behavior is a fundamental characteristic of the current NISQ era, where the circuit depth is strictly bounded by hardware fidelity.

Nevertheless, the robust initial descent of the energy expectation demonstrates the resilience of our non-unitary filtering approach. As quantum hardware matures toward the Early Fault-Tolerant (EFT) regime with improved gate fidelities and partial error mitigation, the performance of our protocol is expected to seamlessly bridge the gap between the noisy results presented here and the ideal theoretical limits established in Sec.~\ref{sec:simulations}.

\section{Derivation of the Non-Unitary Filter via MCMR}
\label{app:filter_derivation}

In this section, we provide a rigorous step-by-step derivation of the non-unitary filter operator $\mathcal{F}$ generated by the mid-circuit measurement and reset (MCMR) protocol on a single ancillary qubit. Let $\mathcal{H}_A$ and $\mathcal{H}_S$ denote the Hilbert spaces of the ancilla qubit and the many-body system, respectively. The initial joint state before the $k$-th filtering step is given by
\begin{equation}
    \ket{\Psi_0} = \ket{0}_A \otimes \ket{\psi^{(k-1)}}_S,
\end{equation}
where $\ket{\psi^{(k-1)}}_S$ is the unnormalized state of the system after $k-1$ successful filtering steps. Applying the Hadamard gate $H_A$ on the ancilla yields a superposition:
\begin{equation}
    \ket{\Psi_1} = (H_A \otimes I_S) \ket{\Psi_0} = \frac{1}{\sqrt{2}} \left( \ket{0}_A + \ket{1}_A \right) \otimes \ket{\psi^{(k-1)}}_S.
\end{equation}
Next, we apply the controlled time-evolution operator $CU(\tau) = \ket{0}\bra{0}_A \otimes I_S + \ket{1}\bra{1}_A \otimes e^{-iH\tau}$. 
This operation entangles the ancilla with the system dynamics:
\begin{align}
    \ket{\Psi_2} &= CU(\tau) \ket{\Psi_1} \\
    &= \frac{1}{\sqrt{2}} \left( \ket{0}_A \otimes \ket{\psi^{(k-1)}}_S + \ket{1}_A \otimes e^{-iH\tau}\ket{\psi^{(k-1)}}_S \right).
\end{align}
A second Hadamard gate is applied to the ancilla to interfere the temporal branches:
\begin{equation}
\begin{aligned}
    \ket{\Psi_3} &= (H_A \otimes I_S) \ket{\Psi_2} \\
    &= \frac{1}{2} \ket{0}_A \otimes \left( I_S + e^{-iH\tau} \right) \ket{\psi^{(k-1)}}_S \\
    &\quad + \frac{1}{2} \ket{1}_A \otimes \left( I_S - e^{-iH\tau} \right) \ket{\psi^{(k-1)}}_S.
\end{aligned}
\end{equation}
Finally, measuring the ancilla qubit in the computational $Z$-basis and post-selecting the outcome corresponding to $\ket{0}_A$ projects the system state into:
\begin{equation}
    \ket{\psi^{(k)}}_S = \bra{0}_A \ket{\Psi_3} = \left( \frac{I_S + e^{-iH\tau}}{2} \right) \ket{\psi^{(k-1)}}_S.
\end{equation}
By cascading this procedure $d$ times via MCMR, the cumulative effect on the initial state $\ket{\psi_{init}}_S$ is exactly the application of the degree-$d$ Power-Cosine filter operator $\mathcal{F}^{(d)} = \left( \frac{I_S + e^{-iH\tau}}{2} \right)^d$.

\section{Robustness to Resonance Detuning}
\label{app:detuning_analysis}

In practical scenarios, the exact ground state energy $E_0$ is generally unknown a priori. 
While coarse estimation techniques can provide an approximate energy $\tilde{E}_0$, it introduces a detuning error $\delta$ such that the resonance condition is imperfectly satisfied. 
In this section, we analyze the robustness of the Power-Cosine filter against such parameter detuning.

Let the applied time step be $\tau = \frac{2\pi}{|\tilde{E}_0|} = \frac{2\pi}{|E_0|(1 + \delta)} \approx \frac{2\pi}{|E_0|} (1 - \delta)$ for a small relative error $\delta \ll 1$. Substituting this imperfect $\tau$ into the eigenvalue transformation function for the ground state, we obtain:
\begin{equation}
    f(E_0) = \cos\left( \frac{E_0 \tau}{2} \right) \approx \cos\left( \pi (1 - \delta) \right) = -\cos(\pi\delta).
\end{equation}
For sufficiently small $\delta$, the Taylor expansion yields $|f(E_0)| \approx 1 - \frac{(\pi\delta)^2}{2}$. 
When the filter of degree $d$ is applied, the unnormalized survival amplitude of the ground state becomes:
\begin{equation}
    |f(E_0)|^d \approx \left( 1 - \frac{\pi^2 \delta^2}{2} \right)^d \approx \exp\left( -\frac{d \pi^2 \delta^2}{2} \right).
\end{equation}
This exponential decay indicates that an inaccurate estimation of $E_0$ suppresses the ground state population alongside the excited states. To guarantee that the ground state is not exponentially heavily damped, the relative detuning error must be bounded by:
\begin{equation}
    \delta \lesssim \mathcal{O}\left( \frac{1}{\sqrt{d}} \right).
\end{equation}
Consequently, as the filter degree $d$ increases to suppress the excited states separated by the spectral gap $\Delta$, the precision requirement for the initial energy estimate $\tilde{E}_0$ becomes more stringent. 
However, since coarse ground state energy estimation algorithms on Early Fault-Tolerant (EFT) devices can achieve Heisenberg-limited precision scaling as $\mathcal{O}(1/T)$, combining an initial low-depth estimation phase with our deterministic state preparation phase forms a theoretically sound and robust pipeline without requiring exact prior knowledge of the spectrum.

\bibliographystyle{elsarticle-num}
\bibliography{bibliography} 

\end{document}